\documentclass[doublecol]{epl2}
\usepackage{graphicx}
\usepackage{color}
\usepackage{amsmath, amsthm, amssymb}
\usepackage[normalem]{ulem}
\newcommand{\be}{\begin{equation}}
\newcommand{\ee}{\end{equation}}
\newcommand{\bea}{\begin{eqnarray}}
\newcommand{\eea}{\end{eqnarray}}

\begin{document}

\title{Phase diagram of a  bidispersed  hard rod lattice gas in two dimensions}
\shorttitle{Bidispersed hard rod gas}
\author{Joyjit Kundu$^1$\and J\"urgen F. Stilck$^2$ \and R. Rajesh$^3$}
\shortauthor{Joyjit Kundu \etal}
\institute{
\inst{1}  Molecular Foundry, Lawrence Berkeley National Laboratory, 1 Cyclotron Road, Berkeley, CA, U.S.A. \\
\inst{2} Instituto de F\'{\i}sica and
National Institute of Science and
Technology for Complex Systems,
Universidade Federal Fluminense,
Av. Litor\^anea s/n,
24210-346 - Niter\'oi, RJ,
Brazil\\
\inst{3} The Institute of Mathematical Sciences, C.I.T. Campus,
Taramani, Chennai 600113, India.}

\date{\today}

\abstract 
{We obtain, using extensive Monte Carlo simulations, virial expansion and a high-density 
perturbation expansion about the fully packed monodispersed phase, the 
phase diagram of a system of bidispersed hard rods on a square lattice.
We show numerically that when the length 
of the longer rods is $7$, two continuous transitions may exist as 
the density of the longer rods in increased, keeping the density of shorter rods fixed: 
first from a low-density isotropic phase to a nematic phase, 
and second from the nematic to a high-density isotropic phase. 
The difference between the critical densities of the two  transitions 
decreases to zero at a critical density of the shorter rods such that the fully packed phase is 
disordered for any composition. When both the rod lengths 
are larger than $6$, we observe the existence of two transitions 
along the fully packed line as the composition is
varied.  Low-density virial expansion, truncated at second virial coefficient, 
reproduces features of the first transition. By developing a high-density perturbation expansion,
we show that when one of the rods is long enough, there will be at least two 
isotropic-nematic transitions along the fully packed line as the  composition is varied.
}


\maketitle

\section{\label{sec:intro}Introduction}

Entropy-driven  transitions in systems of rod-like 
particles have long been an active area of theoretical and experimental 
research. Experimental realizations of such systems include  tobacco 
mosaic virus~\cite{wen1989}, $fd$ 
virus~\cite{eric2008,fraden1997,fraden2000}, silica 
colloids~\cite{jacs2011,kuijk2012}, boehmite 
particles~\cite{buining1993,kooij1996}, DNA origami 
nanoneedles~\cite{nano2014}, liquid crystals~\cite{degennesBook} and 
adsorbed gas molecules on metal 
surfaces~\cite{taylor1985,bak1985,rikvold1991,binder2000,evans2000}.
A system of hard sphero-cylinders in three-dimensional continuum undergoes a 
transition from an  isotropic phase to an orientationally ordered nematic phase 
as density is increased. 
Further increase in density leads to a smectic phase with partial 
translational order and a solid 
phase~\cite{onsager1949,flory1956b,zwanzig1963,frenkel1997,vroege1992,
degennesBook}. In two-dimensional continuum, a Kosterlitz-Thouless 
transition to a high-density phase with power law 
correlations may be observed~\cite{straley1971,frenkel1985,khandkar2005,vink2009}. 
Lattice models of hard rods, of 
interest to this paper, also have a rich phase diagram in two dimensions, 
while not much is known in three dimensions.

Consider monodispersed hard rods on a two-dimensional lattice, 
where each rod occupies $k$ consecutive lattice sites along any of the 
lattice directions and no two rods may overlap. When $k=2$ (dimers), the 
system is known to be disordered at all 
densities~\cite{Heilmann1970,Kunz1970,Gruber1971,lieb1972}. 
When $k \geq 7$, there are, interestingly, two transitions: first, from a low-density 
disordered to an intermediate density nematic phase and second, 
from the nematic to a high-density disordered 
phase~\cite{ghosh2007,joyjit2013}. While the first transition belongs to 
the Ising (three state Potts) universality class for the square 
(triangular) lattice~\cite{fernandez2008a}, the universality class of 
the second transition remains unclear
with the numerically obtained critical 
exponents  differing from those of the first 
transition, though a crossover to the Ising exponents at larger length 
scales could not be ruled out~\cite{joyjit2013,joyjit_rltl2013}. Exact analysis, restricted to a 
rigorous proof for existence of the first transition when
$k \gg 1$~\cite{giuliani2013} and exact solution on a Bethe-like lattice~\cite{dhar2011} 
does not shed any light on the second transition. The fully packed limit of monodispersed
rods is disordered and
may be mapped onto a height model with a $k-1$ dimensional 
height field, showing that 
orientation-orientation correlations decay as a power law~\cite{lieb1972,henley1997,ghosh_jesper2007}.

Polydispersity in length of the particles is hardly avoidable in 
experiments and results in features such as strong fractionation, 
two distinct nematic phases and  nematic-nematic or 
isotropic-nematic-nematic phase coexistence~\cite{buining1993, 
kooij1996}. Some of these features may be obtained using density
functional theory, virial expansion or 
Monte Carlo simulations in the continuum~\cite{coulon1984,Birshtein1988,vroege1993,varga2000,
sollich2003,Speranza2003,bohle1996,frenkel1998,cuesta2000,cuesta2003}. 
The lattice counterpart is less studied and the
phase diagram is  mostly unexplored. A particular 
model of polydispersed rods with a rod of length $k$ having a weight $z_e^2 
z_i^{k-2}$, where $z_i$ ($z_e$) is the fugacity of an internal 
(endpoint) monomer was shown to undergo an isotropic-nematic transition 
using transfer matrix methods~\cite{jurgen2015}.  When 
$z_e=\sqrt{z_i/2}$, the model may solved exactly by mapping it to the 
two-dimensional Ising model~\cite{velenik2006}.  A second transition to the high-density disordered 
phase is absent~\cite{jurgen2015}. However, in this model, densities of different
species cannot be changed independently.

What is the phase diagram for lattice models of polydispersed rods? Does polydispersity preserve the 
second phase transition into a high-density disordered phase? Is the fully packed 
line still  disordered or could there
be regions with nematic order? In this letter, we address these questions
by determining the phase diagram of bidispersed 2-7, 6-7, and 7-8 mixtures using
extensive Monte Carlo simulations and  studying generic bidispersed mixtures
using low-density virial expansions and high-density perturbation expansions close to
full packing. In particular, we show that the second transition at high densities persists, and if one
of the rod lengths is large enough, the system at full packing will exhibit at least two
isotropic-nematic  transitions as the ratio of 
densities of the two species is varied.

\section{\label{sec:model_ch01} Model and the Monte Carlo algorithm}

Consider a bidispersed system of rods of length $k_1$ and $k_2$ on a 
square lattice of size $V=L\times L$ with periodic boundary conditions, 
where each rod is either horizontal or vertical.  A horizontal 
(vertical) rod of length $k_i$ (where $i = 1,2$) occupies $k_i$ 
consecutive lattice sites along the $x$ ($y$)-axis. No two rods are 
allowed to intersect or equivalently, each site may be occupied by 
utmost one rod. A fugacity $z_{k_i}=e^{\mu_{k_i}}$ is associated with 
each rod of length $k_i$, $i=1,2$, where $\mu_{k_i}$ is the 
corresponding reduced chemical potential.

We simulate this model using a constant fugacity grand canonical Monte 
Carlo algorithm involving cluster moves. For fixed fugacities $z_{k_1}$ and 
$z_{k_2}$, the system reaches an equilibrium density $\rho (z_{k_1},z_{k_2})$, defined as the fraction of 
sites occupied by the rods. This algorithm is an adaptation 
of the scheme that was quite efficient in equilibrating systems of 
monodispersed long rods~\cite{joyjit_dae,joyjit2013}. Variants of 
this algorithm have been used to study systems of hard 
rectangles~\cite{joyjit_rectangle,joyjit_rectangle_odd,joyjit_rec_asym}, 
disks on square lattice~\cite{trisha_knn} and mixtures of 
squares and dimers~\cite{kabir2015}. We briefly discuss the algorithm 
here.

Choose at random a row or column of the lattice. If a row is chosen, all the horizontal 
rods on that row are removed, while the rest of the configuration is kept unchanged. 
The row now consists of intervals of empty sites separated by the sites occupied by 
vertical rods. These empty intervals are re-occupied with a new configuration 
of horizontal rods consistent with equilibrium probabilities. If, instead of a row, a 
column is chosen, a  similar evaporation-deposition operation is done with vertical 
rods. The calculation of these equilibrium probabilities reduces to a one-dimensional 
problem. 

Let $\Omega_o(z_{k_1},z_{k_2};\ell)$ be the grand canonical partition 
function of a one-dimensional chain of length $\ell$ with open boundary 
conditions. The probability that the first site of the one-dimensional 
chain is occupied by the left-most site of a rod of length $k_i$ is 
$p^{\ell}_i= z_i 
\Omega_o(z_{k_1},z_{k_2};\ell-k_i)/\Omega_o(z_{k_1},z_{k_2};\ell)$, 
where $i=1, 2$. The partition functions $\Omega_o(z_{k_1},z_{k_2};\ell)$ 
obeys the recursion relation $\Omega_o(z_{k_1},z_{k_2};\ell)= 
\sum_{i=1}^2 z_{k_i} \Omega_o(z_{k_1},z_{k_2};\ell-k_i)+ 
\Omega_o(z_{k_1},z_{k_2};\ell-1)$ for $\ell \geq \min(k_1,k_2)$, with 
the boundary conditions $\Omega_o(z_{k_1},z_{k_2};\ell)=1$ for 
$\ell=0,1,\dots, \min(k_1,k_2)-1$ and $\Omega_o(z_{k_1},z_{k_2};\ell)=0$ 
for $\ell<0$. The partition function of a one-dimensional chain of 
length $\ell$ with periodic boundary condition, 
$\Omega_p(z_{k_1},z_{k_2};\ell)$, is easy to determine once 
$\Omega_o(z_{k_1},z_{k_2};\ell)$ is known. It obeys the recursion 
relation $\Omega_p(z_{k_1},z_{k_2};\ell)= \sum_{i=1}^2 z_{k_i} k_i 
\Omega_o(z_{k_1},z_{k_2};\ell-k_i)+ \Omega_o(\ell-1)$. The recursion 
relations may be solved exactly for $\Omega_o(z_{k_1},z_{k_2};\ell)$ and 
$\Omega_p(z_{k_1},z_{k_2};\ell)$. The list of relevant probabilities 
${p^\ell_i}$ for all $\ell \leq L$ are stored in order to reduce 
computational time.

In addition to the evaporation-deposition moves, we also implement a 
flip move~\cite{joyjit_rectangle}. We choose a site at random. Only if 
it is the bottom-left corner of a block of size $(k_i \times k_i)$ 
containing $k_i$ aligned parallel horizontal (vertical) rods, it is 
replaced by a similar block of $k_i$ vertical (horizontal) rods. One 
Monte Carlo (MC) move contains $2L$ evaporation-deposition moves and $L^2$ 
flip moves. All the numerical results presented in this paper are 
obtained using a parallelized version of the algorithm.

The largest system size that we simulate is $L=560$.  A single data point in
the phase diagrams (total of 47 data points), has been obtained using 
(on an average) 30 runs of Monte Carlo simulation.
\begin{figure*} 
\begin{center} 
\includegraphics[width=1.6\columnwidth]{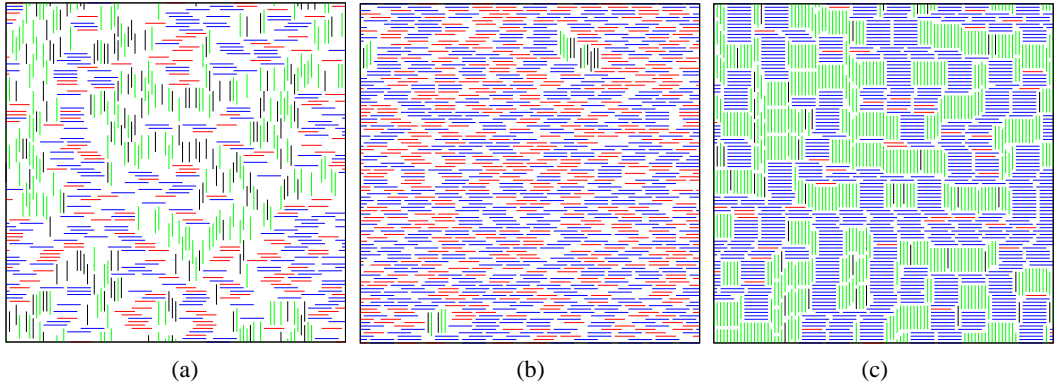}
\caption{Snapshots of the system when $k_1=7$ and $k_2=8$ at (a) low-density 
isotropic (I) phase where $\rho_7 \approx 0.187$ and 
$\rho_8\approx 0.311$ ($\mu_7=-2.0$, $\mu_8=-1.4$), (b) intermediate 
density nematic (N) phase where $\rho_7\approx 0.372$ and $\rho_8\approx 
0.441$ ($\mu_7=1.5$, $\mu_8=2.0$) and (c) high-density isotropic (I) 
phase where $\rho_7\approx 0.055$ and $\rho_8\approx 0.922$ 
($\mu_7=7.2$, $\mu_8=11.0$). The horizontal and vertical rods of length 
$7$ are colored red and black, and the same of length $8$ are colored 
blue and green respectively.}
\label{fig:snap}
\end{center}
\end{figure*}

\section{\label{sec:results}Results}

We study three different mixtures: 2-7, 6-7, and 7-8. These choices 
were made for the following reasons. A monodispersed system of hard 
rods shows phase transition only when the rod length $k \geq 7$. Rods of 
length $2$ and $6$ being the smallest and largest lengths that do not 
show a nematic phase, studying 2-7 and 6-7 allows us to obtain the 
trend for intermediate lengths. To study the effect of mixing rods of 
different lengths, both of which show nematic phase, we study the 7-8 
mixture. We observe three phases: a low-density isotropic (I) phase,
a nematic (N) phase and a high-density
isotropic (I) phase. Typical snapshots of these phases for the 7-8  mixture are shown in  Fig.~\ref{fig:snap},
where $\rho_{k_1}$ and 
$\rho_{k_2}$ denote the densities (fraction of occupied sites) of the two species.

\subsection{\label{sec:phase_diagram}Phase diagrams}

\begin{figure*} 
\begin{center} 
\includegraphics[width=1.7\columnwidth]{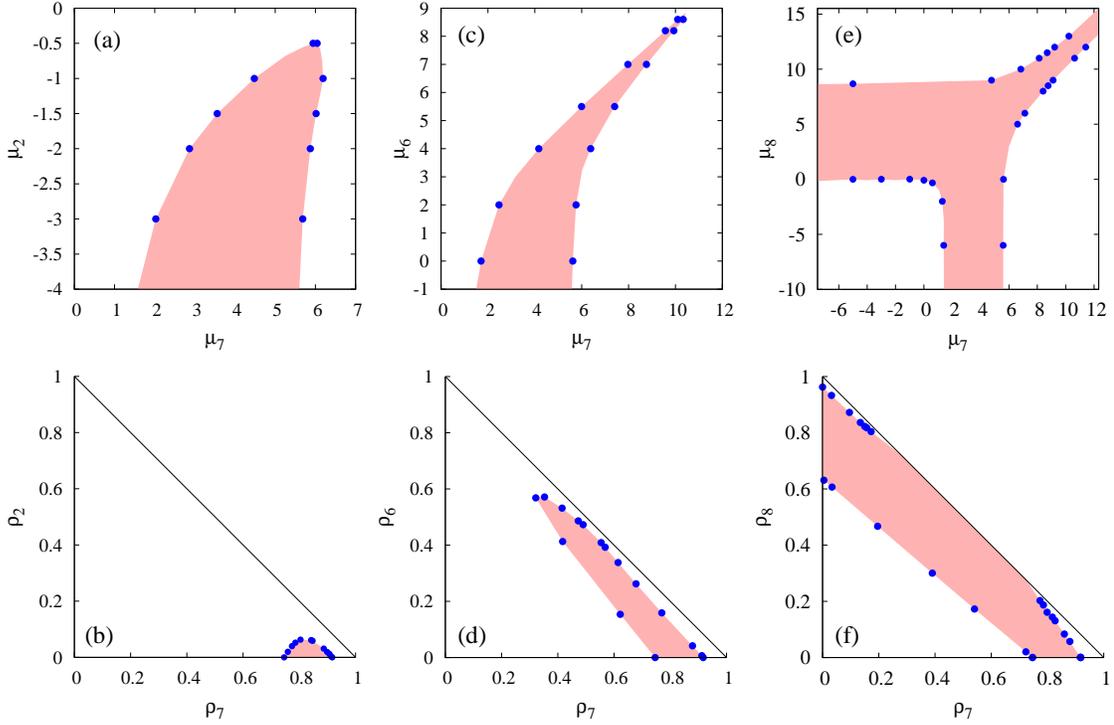}
\caption{Phase diagram in the $\mu$-plane for mixtures of (a) $2$ and $7$, 
(c) $6$ and $7$, and (e) $7$ and $8$, and in the $\rho$-plane for 
(b) $2$ and $7$, (d) $6$ and $7$, and (f) $7$ and $8$. The data 
points (solid circles) are obtained from Monte Carlo simulations. The 
shaded regions are guides to the eye and correspond to regions with 
nonzero nematic order. }
\label{fig:phase_diag}
\end{center}
\end{figure*}

For a particular bidispersed mixture, we obtain the complete phase 
diagram by simulating the system at different values of $\mu_{k_1}$ and 
$\mu_{k_2}$. The critical chemical potentials and densities are 
determined from the crossing of the curves of the 
Binder cumulant as a 
function of the chemical potential or density for different system sizes. 
The phase diagrams  for 2-7, 6-7, and 7-8 mixtures are 
shown in Fig.~\ref{fig:phase_diag}. The shaded regions in the phase 
diagrams correspond to ordered N phases, while the empty 
regions correspond to I phases with no orientational order.

A system of monodispersed dimers ($k=2$) does not show any phase 
transition. Thus, when $\rho_2 \gg \rho_7$ (for 2-7 mixture), we do not expect any phase 
transition. When $\mu_2$ or $\rho_2$ is small enough, we 
observe two transitions as $\mu_7$ or $\rho_7$ is increased: first from 
a low-density I phase to an intermediate density N phase and second from 
the N phase to a high-density I phase [see Fig~\ref{fig:phase_diag}(a) 
and (b)], as seen for the system of monodispersed rods of length $\geq 7$. 
The difference between the two critical densities decreases as 
$\rho_2$ is increased and beyond a critical $\rho_2$, no transitions are observed. 
On the other hand, when $\rho_7$ is kept fixed and $\rho_2$ is increased,
utmost one transition is present. One may go from the low-density I phase to the 
high-density I phase continuously without crossing 
any phase boundary, suggesting that the high-density I phase is a
re-entrant low-density I phase~\cite{joyjit_rltl2013}. 

The phase diagram for 6-7 mixture [see Fig.~\ref{fig:phase_diag}(c) and (d)] 
is quite similar to that of the  2-7 mixture.  The area of the 
nematic region is larger for the 6-7 mixture,
showing that longer rods favor orientational ordering. Unlike
the 2-7 mixture, now there are regions where the system undergoes two transitions 
when $\rho_7$ is kept fixed and $\rho_6$ is varied. The fully
packed line remains disordered for all compositions of 2-7 and 6-7 mixtures. 
We  expect a qualitatively similar phase diagram for mixtures with $k_1<7$ and $k_2=7$.

Now, consider the 7-8 mixture. This case is different from the above
two, as there are two critical points on each of the two axes $\rho_7=0$ and  $\rho_8=0$
[see 
Fig.~\ref{fig:phase_diag}(e) and (f)]. For small values
of $\rho_7$ or $\rho_8$, two transitions are observed, The 
high-density I phase is separated
from the low-density I phase by a region of N phase.
It raises the 
question whether the system is disordered at full packing as seen for
2-7 and 6-7 mixtures. The algorithm that we use does not
equilibrate the system at full packing. Instead, by simulating the system
close to full packing,  we find that the phase 
boundaries, separating the N phase from the high-density I phase
approach the $\rho=1$ line, and appear to terminate at two separate 
points [see Fig.~\ref{fig:phase_diag}(f)], suggesting that there are two transitions
along the $\rho=1$ line.

All the transitions that we observe are continuous as we do not observe any 
jump in the density or in the nematic order parameter near the transitions. Also, the 
nematic order of both the species increases from zero simultaneously with 
density, and thus we do not observe any fractionation effect.

\subsection{\label{sec:virial}Virial Expansion}

We now determine the phase diagram of the system from a  
standard low-density virial expansion for multiple species
truncated at the second virial coefficient~\cite{zwanzig1963}.

Let $N_{k_i}^j$, where $i=1,2$ and $j=h$ (horizontal), $v$ (vertical), 
denote the number of rods of length $k_i$ with orientation 
$j$. The partition function of 
a system of $N$ rods in a  volume $V$ is then given by
\be
Q_N =\frac{V^N}{N! 4^N} \sum_{\{ N_{k_i^j}\}}'
\frac{N!}
{\prod_{i,j}N_{k_i}^j!}
\exp[-\phi(\{N_{k_i}^j\})],
 \label{eq:final_partition}
\ee
where the prime denotes the constraint 
$\sum_{i,j} N_{k_i}^j=N$, and $\phi_N$ is the reduced excess free
energy for a given distribution of lengths and orientations:
\be
\exp[-\phi_N (\{N_{k_i}^j\} )]=\frac{1}{V^N}  \sum_\textbf{R} 
\exp(-\beta U_N),
\label{eq:excess_free}
\ee
where $U_N$ is the total interaction energy, and
$\textbf{R}$ denotes all possible positions.

Let $x_i^j=N_{k_i}^j/N$ denote the fraction of rods of length
$k_i$ with orientation $j$. For large 
$N$, $V$,   \eqref{eq:final_partition} may be
written as
\be
Q_N = \int_0^1 \prod_{i,j} d x_i^j e^{-N F(\{x_i^j\})}
\delta\left(\sum_{i,j} x_i^j-1\right),
\label{eq:partition_sum}
\ee
where $F$ is the free energy per particle: 
\be
F(\{x_i^j\})=\ln\frac{4 \theta}{e} + \sum_{i,j}
x_i^j \ln x_i^j 
+\frac{1}{N}\phi_N(\{x_i^j\}), \label{eq:f_rho}
\ee
and $\theta$ is the total number density of rods.
For large $N$, the integrals in \eqref{eq:partition_sum} may be replaced by the
largest value of the integrand with negligible error. Thus, the values
of $x_i^j$ are determined by minimizing the free energy in \eqref{eq:f_rho}.

We compute the reduced excess free energy $\phi_N$ as a virial
expansion. For a 
composition $\textbf{x}=(x_1^h,x_1^v,x_2^h,x_2^v)$, the expansion is 
\be
-\frac{1}{N} \phi_N(\theta, \textbf{x})= \sum_{n=2} 
B_n(\textbf{x}) \theta^{n-1}, \label{eq:virial}
\ee
where $B_n$ is the $n$-th virial coefficient:
\be
B_n(\textbf{x})=\frac{1}{V n!} \sum'_{\{n_{k_i}^j\}}
\frac{n!}{\prod_{ij} n_{k_i}^j}
\prod_{ij}(x_i^j)^{n_{k_i}^j} 
B(\{n_{k_i}^j\}). 
\label{eq:def_vir}
\ee
Here, $n_{k_i}^j$ is the number of rods of length $k_i$ and
orientation $j$ in a irreducible graph of size $n$, 
the prime denotes the constraint 
$\sum_{i,j} n_{k_i}^j=n$, and 
$B(\{n_{k_i}^j\})=\int \sum \prod f$ is the standard 
abbreviation for the cluster integrals having Mayer functions $f$ over the irreducible graphs 
consisting of $n_{k_i}^j$ rods of length $k_i$ and orientation $j$. 

We truncate the expansion in \eqref{eq:virial} at the second virial 
coefficient.  On a lattice, the evaluation of 
the virial coefficients reduces to the problem of counting the number of 
disallowed configurations. We thus obtain 
$B(2,0,0,0)=B(0,2,0,0)=-V (2 k_1-1)$, 
$B(0,0,2,0)=B(0,0,0,2)=-V (2 k_2-1)$, 
$B(1,0,0,1)=B(0,1,1,0)=-V k_1 k_2$, 
$B(1,0,1,0)=B(0,1,0,1)=-V (k_1+k_2-1)$, 
$B(1,1,0,0)=-V k_1^2$, $B(0,0,1,1)=-V k_2^2$.
On substituting the virial coefficients into 
\eqref{eq:def_vir}, \eqref{eq:f_rho} reduces to
\bea
F(\textbf{x}) &=& \ln \frac{4\theta}{e} + \sum_{ij}x_i^j \ln x_i^j 
+\frac{\theta}{2} \sum_{ij} (x_i^j)^2 (2 k_i-1) \nonumber \\
&+& \theta \sum_i k_i^2 \prod_j x_i^j 
+ \theta  (k_1+k_2-1) \sum_j  \prod_i x_i^j 
\nonumber \\
&+& \theta (x_1^h x_2^v+x_2^h x_1^v) k_1 k_2+O(\theta^2).
\label{eq:freeenergy_final}
\eea

For given densities of the two species, the free energy $F$ in
\eqref{eq:freeenergy_final} may be expressed in terms of 
the nematic order parameters of the two species 
denoted by $\psi_{k_1}=(x_1^h-x_1^v)/(x_1^h+x_1^v)$ 
and $\psi_{k_2}=(x_2^h-x_2^v)/(x_2^h+x_2^v)$. 
The phase for a given number density is obtained by minimizing $F$ 
with respect to $\psi_{k_1}$ and $\psi_{k_2}$. 
We find existence of only the low-density isotropic-nematic (I-N)
transition,  similar to  the
solution of the monodispersed system on a Bethe-like lattice~\cite{dhar2011}. 
The I-N phase boundary may be obtained by solving the equation 
$\frac{\partial^2 F}{\partial^2 \psi_{k_1}} \frac{\partial^2 F}{\partial^2 \psi_{k_2}}-
\left(\frac{\partial^2 F}{\partial \psi_{k_1}\partial \psi_{k_2}}\right )^2=0$, and we obtain 
\be
\frac{(k_1-1)^2}{k_1} \rho_{k_1}^c + \frac{(k_2-1)^2}{k_2} \rho_{k_2}^c=2.
\label{eq:critical}
\ee
By setting $\rho_{k_2}^c=0$, we obtain 
the critical density $\rho_{k_1}^c= 2k_1/(k_1-1)^2$, as found earlier for
the  monodispersed system~\cite{joyjit_rec_asym}. For the monodispersed
system, the I-N transition exists for lengths larger than $3$.

\begin{figure}
\includegraphics[width=\columnwidth]{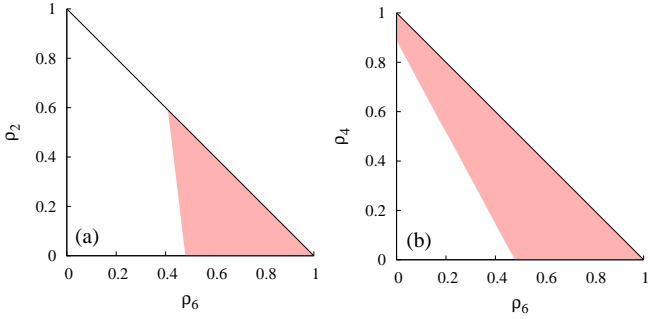}
\caption{Phase diagram in the $\rho$-plane for mixtures of (a) $2$ and $6$,
and  (b) $4$ and $6$, 
obtained from virial expansion. The shaded regions correspond to N phase 
and the empty regions correspond to  I phase.}
\label{fig:vir_phase}
\end{figure}

The phase diagrams for two different mixtures (2-6 and 4-6), obtained
from virial expansion are shown in 
Fig.~\ref{fig:vir_phase}. The shaded (empty) regions correspond to
N (I)  phases. 
While the theory predicts the 
existence of an I-N transition at full packing ($\rho=1$) for 2-4 mixture as 
the ratio of the densities of the two species are varied [see  
Fig.~\ref{fig:vir_phase}(a)], for the 4-6 mixture, the fully packed line is always nematic [see  
Fig.~\ref{fig:vir_phase}(b)]. Within the virial theory,
we do not observe any 
fractionation effect or equivalently, the nematic order for both the 
rods increases from zero simultaneously as the total density is varied.

The phase diagram obtained from virial expansion
differ from those obtained by simulations. 
It was shown in Ref.~\cite{joyjit_rec_asym} that in two 
dimensions, the higher order virial coefficients do contribute and can 
not be neglected even in the limit $k_i \to \infty$. Thus, truncating 
the expansion of the reduced excess free energy at the second virial 
coefficient is only a reasonable approximation at very low densities.

\subsection{\label{sec:fullpacked}Expansion about the pure state along the fully packed line}

The fully packed line can neither be numerically studied with the algorithm 
used in this paper nor with the low-density virial expansion. Instead, we calculate the entropies of the 
I and N phases as a perturbation expansion about the 
fully packed monodispersed state. For simplicity, let $k_1$ and $k_2$ 
be mutually prime. We approximate the N phase as one where all the 
rods point in one direction. The arrangement 
of these rods is a simple combinatorial problem and the entropy per unit 
site $s_{nem}$, in terms of the number densities of the two species $\theta_{k_1}$ and 
$\theta_{k_2}$, is
\be
s_{nem} \approx (\theta_{k_1} + \theta_{k_2}) \ln(\theta_{k_1} + \theta_{k_2})-\theta_{k_1} \ln 
\theta_{k_1} - \theta_{k_2} \ln \theta_{k_2},
\label{eq:snem}
\ee
where $k_1 \theta_{k_1} + k_2 \theta_{k_2}=1$. Expanding 
\eqref{eq:snem} for 
small $\theta_{k_2}$, we obtain
\be
s_{nem} \approx - \theta_{k_2} \ln \theta_{k_2} + O(\theta_{k_2} ), ~ \theta_{k_2} \to 0.
\label{eq:snem_asy}
\ee

To estimate the entropy of the I phase, we break the lattice 
into $L/k_1$ horizontal strips of width $k_1$. The partition function ${\mathcal 
L}_0$, when only rods of length $k_1$ are present is then,
\be
{\mathcal L}_0 = 2 k_1 \omega_p(L)^{L/k_1},
\ee
where $\omega_p(L)$ [$\omega_o(L)$] is the partition function for a 
strip of length $L$ with periodic [open] boundary conditions, and the 
factor $2 k_1$ accounts for the two orientations and translational 
invariance. Clearly, $\omega_o(L) = \omega_o(L-1) + \omega_o(L-k_1) $ 
with solution $\omega_o(L) = a_o \lambda^L$, where 
$\lambda^{k_1}-\lambda^{k_1-1}-1=0$. Likewise, $\omega_p(L) = a_p 
\lambda^L$.

\begin{figure}
\begin{center}
\includegraphics[width=3.5cm]{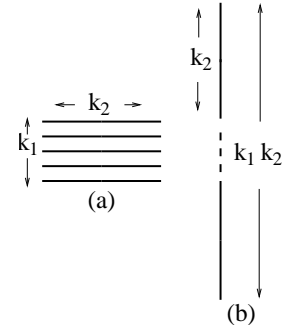}
\caption{Defects, made up of rods of length $k_2$, that contribute at the lowest order
in the perturbation expansion about the fully packed case with only rods of length $k_1$.}
\label{fig:expansion}
\end{center}
\end{figure}

Consider defects consisting of rods of length $k_2$. To make the system 
fully packed, a minimum of $k_1$ such rods are required. If $k_1$ and 
$k_2$ were not mutually prime, this number would change. The smallest 
contribution to the partition function is when these rods are arranged 
in a plaquette of size $k_1\times k_2$ as 
in Fig.~\ref{fig:expansion}(a) or in a vertical line as in Fig.~\ref{fig:expansion}(b). 
Denoting the contribution from these defects as 
${\mathcal L}_1$, we obtain
\be
{\mathcal L}_1 =2 L^2 z_{k_2}^{k_1}  \omega_p(L)^{L/k_1} \left[ 
\frac{\omega_o (L-k_2) }{ \omega_p(L)}
+   \frac{\omega_o (L-1)^{k_2}}{\omega_p(L)^{k_2}} \right].
\ee

Substituting for $\omega_p$ and $\omega_o$ in terms of $\lambda$, we obtain the 
partition function ${\mathcal L}$ to be
\bea
{\mathcal L} &=& 2 k_1 \left[a_p \lambda^L\right]^{L/k_1} \times 
\nonumber \\
&&\left[1+\frac{L^2 z_{k_2}^{k_1}} {k_1 \lambda^{k_2}} 
\left[\frac{a}{a_p} + 
\left(\frac{a}{a_p}\right)^{k_2}\right]+O\left(z_{k_2}^{2 k_1}. 
\right)\right]
\eea
The free energy, $-\ln {\mathcal L}$, is in terms of the fugacity. Performing
a Legendre transform to obtain entropy in terms of density, we find
\be
s_{iso}\approx \frac{\ln \lambda + n_{k_2}}{k_1}-\frac{n_{k_2}}{k_1} \ln 
\left[\frac{n_{k_2} 
\lambda^{k_2}}{\frac{a}{a_p}+\left(\frac{a}{a_p}\right)^{k_2} } \right] 
+ \ldots.
\label{eq:disordered}
\ee

For large $k_1$, the calculation based on strips gives a good estimation 
of the entropy. In this limit, $\ln \lambda \approx k_1^{-1} \ln k_1$~\cite{ghosh2007}. 
Equating the entropies for N and I phases [see \eqref{eq:snem_asy} and \eqref{eq:disordered}], 
we obtain that along the fully packed line, 
the system undergoes a isotropic-nematic transition at $\theta_{k_2}^c \sim 
k_1^{-2}$. Given that the fully packed phase of monodispersed system
is isotropic, we expect that there are at least two I-N transitions
along the fully packed line,  when $k_1$ is very large.

\section{Discussions}

In this letter, we determined the phase diagram of a system of bidispersed hard rods 
on a square lattice using Monte Carlo simulations, virial expansion and high-density 
perturbation expansion. Numerically, the phase diagrams of  three different mixtures (2--7, 6--7 and 7--8)
were determined.
For any $2\leq k_1 \leq 6$ and $k_2=7$, the system at full packing is 
always disordered and the phase diagram is expected to be 
qualitatively similar to that of 6--7 or 2--7 mixture. When $k_1, k_2 \geq 7$, we expect 
the phase behavior to be qualitatively similar to that of 7--8 mixture, and predict the existence of 
two transitions at full packing. The low-density  virial expansion is able to reproduce the low-density
I-N transition but does not work well at high densities. When one of the rod lengths is high enough, 
the high-density perturbation expansion predicts the existence of two I-N phase 
transitions along the fully packed line. This prediction could not be
verified numerically as the algorithm used in the paper is unsuitable for studying the fully packed line. 
 
Monodispersed hard rectangles have a richer 
phase diagram than rods, with up to three density driven transitions: 
from isotropic to nematic to columnar to a solid-like 
phase~\cite{joyjit_rectangle,joyjit_rectangle_odd,joyjit_rec_asym,nkr14}. A simple
mixture of dimers and  squares shows a line of critical points with continuously varying
exponents~\cite{kabir2015}. Polydispersed
rectangles are thus expected to have a complicated phase diagram and in three dimensions may 
show features like fractionation, and 
therefore is a promising area for future study.

\begin{acknowledgments}
The simulations were done on the supercomputer
Annapurna at the Institute of Mathematical Sciences. 
\end{acknowledgments}


\end{document}